\documentclass{iau} 
\usepackage{graphicx}
\usepackage{subfig}
\usepackage[font=small,skip=0pt]{caption}

\title[Spatial Inhomogeneity In Solar Faculae] {Spatial Inhomogeneity In Solar Faculae}

\author[A. Elek et al.] {A. Elek$^{1,3}$, N. Gyenge$^{1,2}$, M. B. Kors\'os$^{2,1}$, R. Erd\'elyi$^{2,3}$}

\affiliation{$^1$Debrecen Heliophysical Observatory (DHO), Konkoly Observatory,\\
Research Centre for Astronomy and Earth Sciences\\ 
Hungarian Academy of Sciences, Debrecen, P.O.Box 30, H-4010, Hungary
\\email: {\tt elek.anett@csfk.mta.hu} \\[\affilskip]
$^2$Solar Physics and Space Plasmas Research Centre (SP$^{2}$RC),\\
University of Sheffield, Hicks Building, Hounsfield Road, Sheffield S3 7RH, England (UK)\\
$^3$Dept. of Astronomy, E\"otv\"os L. University, P\'azm\'any P. s\'et\'any 1/A, Budapest, H1117,Hungary}

\pubyear{2017}
\volume{335}
\jname{Space Weather of the Heliosphere: Processes and Forecasts}
\editors{C. Foullon \& O. Malandraki}

\begin{document}

\maketitle

\begin{abstract}

 In this paper, we investigate the inhomogeneous spatial distribution of solar faculae. The focus is on the latitudinal and longitudinal distributions of these highly localised features covering ubiquitously the solar surface. The statistical analysis is based on white light observations of the Solar and Heliospheric Observatory (SOHO) and Solar Dynamics Observatory (SDO) between 1996 and 2014. We found that the fine structure of the latitudinal distribution of faculae displays a quasi-biennial oscillatory pattern. Furthermore, the longitudinal distribution of photospheric solar faculae does not show homogeneous behaviour  either. In particular, the non-axisymmetric behaviour of these events show similar properties as that of the active longitude (AL) found in the distribution of sunspots. Our results, preliminary though, may provide a valuable observational constrain for developing the next-generation solar dynamo model.

%\keywords{Sun: activity, Sun: faculae, plages}

\end{abstract}

\firstsection
\section{Motivation}

The connection of the small-scale magnetic field of solar photospheric faculae and the global magnetic field is studied, e.g., in the context of solar cycle dependence (\cite{Ermolli2014, chatterjee2016}) or the latitudinal distribution \cite{Makarov1994}. These studies highlight an important question, as it is not obvious, whether photospheric faculae should follow the global characteristics of active regions (ARs). ARs, as widely accepted, are the surface appearance of the solar global dynamo operating at the tachocline at the bottom of the convection zone. Global dynamo models work relatively well to describe the measured low-latitude global magnetic field. However, such global dynamos may struggle to adequately capture the physics of high-latitude magnetic phenomena, not to mention those that appear outside of ARs, or magnetic phenomena that may be generated by local dynamo, close to the solar surface. Since it is unclear whether the local dynamo itself indeed exists, it is even more unclear whether there is an interaction of the global and local dynamos. Therefore, studying the statistical properties of the highly localised photospheric  faculae on spatial and temporal time scales well beyond their own extent and lifetime may provide further observational constrains to dynamo theorems. Valid and very interesting questions that should be answered are, such as e.g. (i) do faculae show indeed the butterfly distribution; (ii) how to account for the high-latitude distribution of faculae \cite{Deng2016}; (iii) is there a longitudinal pattern, supporting the mounting evidence of active longitudes (AL), of solar faculae; (iv) are all faculae the result of the global 11-year cycle dynamo, or, could there be a second dynamo operating (e.g. a local, surface one); (v) could actually the statistical analysis of spatio-temporal properties of faculae provide evidence towards the coupling of global and local dynamos; - just to raise a few questions that may justify why to pay so much attention to faculea. 

\begin{figure}
\centering
\includegraphics[width=1\linewidth]{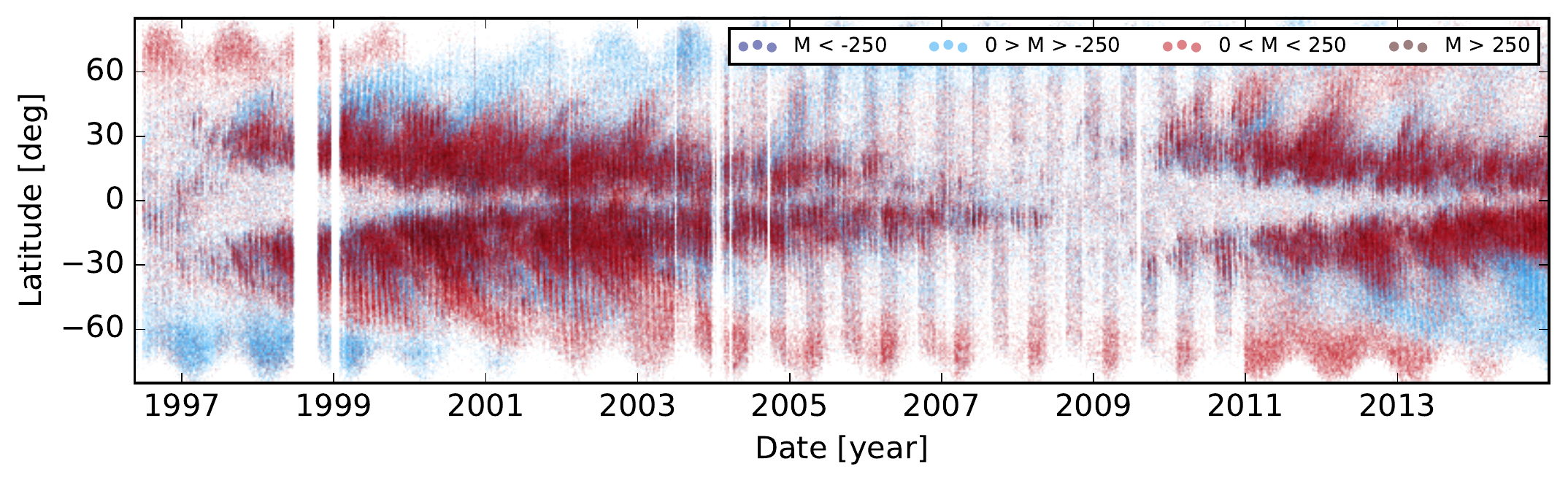}
\caption{The latitudinal distribution of solar faculae based on SOHO and SDO observations. The dot colours refer to the associated magnitude of magnetic field of a given facula. Features, where $A < 2$, are omitted from the SDO data.}
\label{fig1}
\end{figure}

\section{Solar Faculae Catalogue}

We used the solar faculae database by the Debrecen Heliophysical Observatory (for details see \cite{Baranyi2016} and \cite{Gyori2017}). The wightlight observations are provided by the Solar and Heliospheric Observatory (SOHO) and Solar Dynamics Observatory (SDO) satellites. The catalogue contains information about the date of a facula event, area $A$ in millions of solar hemisphere, position in Carrington coordinate system (latitude $B$ and longitude $L$) and magnetic field strength ($M$). The database provides hourly cadence between 1996 and 2014. Our statistical sample includes all the observed solar faculae, except those data which failed one of the following criteria: (i) the longitudinal distance from the central meridian must be closer than 75 degrees and (ii) the absolute magnetic field strength must be less than 500 G. Both criteria are introduced due the errors in magnetograms. We distinguish four sub-samples based on the magnetic field strength (i) $M$: $M > 250$, (ii) $0 < M < 250$, (iii) $0 > M > -250$ and (iv) $M > -250$.

\section{Latitudinal and Longitudinal Distribution of Solar Faculae}

Figure \ref{fig1} displays the latitude-time diagram of our statistical sample. The general shape of this diagram is the well-known butterfly-diagram. By following the global latitudinal migration towards the equator there seems to be a polar ward migration in narrow bands that repeats itself. Figure \ref{fig1} shows that this migration is not always parallel, resulting in varying periodicity as function of latitude. The fine structure of the distribution reveals itself the negative and positive weak magnetic fields ($\left| B \right| < 250$) alternate. 

\begin{figure}
  \centering
  \includegraphics[width=1\linewidth]{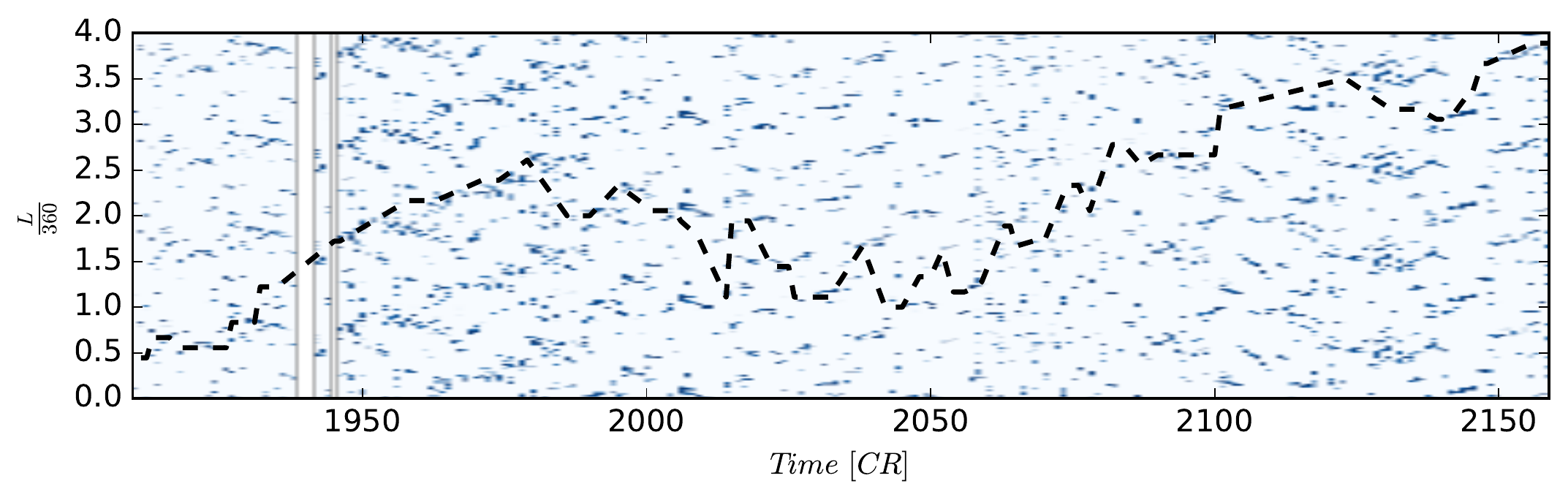}
  \vspace{0mm}
  \includegraphics[width=1\linewidth]{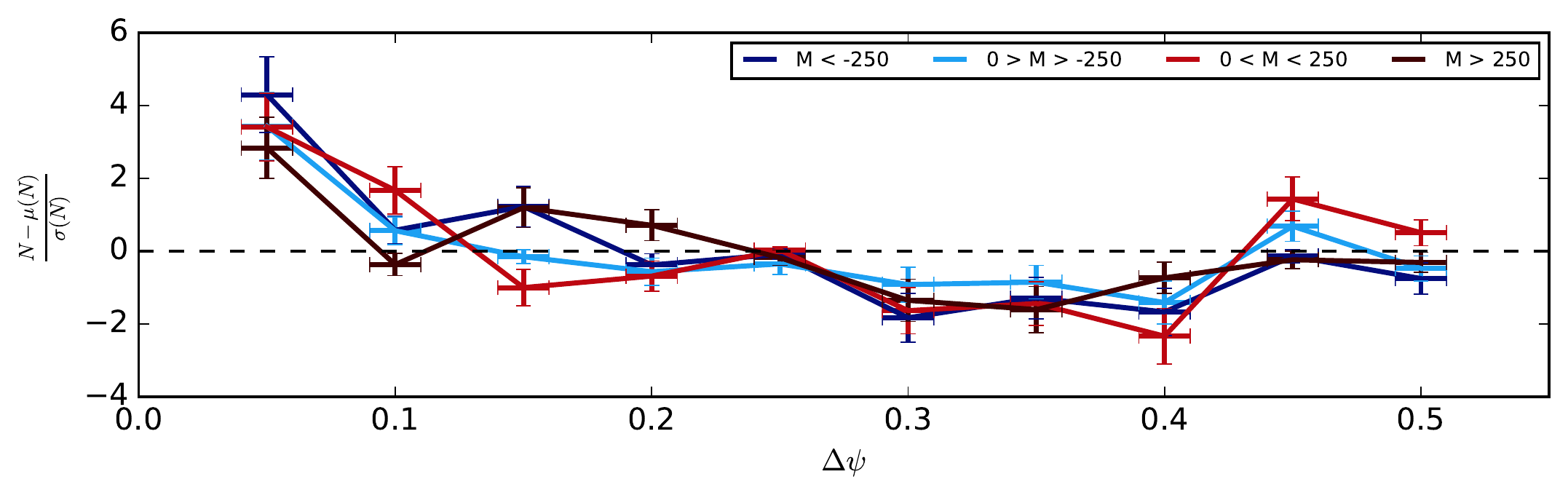}
\caption{Upper panel: The longitudinal distribution of solar faculae. The dashed line indicates the migration of AL based on sunspot groups. Bottom panel: The distance of the AL migration and the solar faculae. The coloured lines represent the magnetic field strength. The error bars in $x$, due the differential rotation, while $y$ are given by the square root of the certain bin.}
\label{fig2}
\end{figure}

The upper panel of Figure \ref{fig2} shows the longitudinal distribution of solar faculae while the bottom panel plot shows the faculae longitudinal distance from the AL. The position of AL is defined in \cite{Gyenge2017}. Essentially, the AL is the most enhanced longitudinal activity in a certain Carrington rotation, based on sunspot data, i.e the longitudinal distribution of sunspot groups is non-homogeneous. The parameter $\delta\phi$ represents the distance between the AL and the faculae in Carrington Phases. If $\delta\phi = 0$, the longitudinal position of the faculae is same as the longitudinal position of the AL, meanwhile $\delta\phi = 0.5$ corresponds to a 180 degrees shift. We summarised the faculae in each $\delta\phi = 0.05$ width bins. The obtained total counts $N$ are converted to standard deviation units $(N-\mu(N)) / \sigma(N)$, displayed by the vertical axis. 

We conclude that he latitudinal migration shows the presence of the 11-year solar cycle. The fine structure also reveals a quasi-biennial oscillation - not highlighted in the displayed figures, however, a time series analysis reveals it. We determined this oscillation as follows: we cut the sunspot latitude-time diagram into 5-degree segments and analysed the temporal variation histograms of the total magnetic field based on each segments. These histograms show periodic recurrent features with 1-2 years oscillation period. The periodicity may be dependent on latitude as the narrow brands are not parallel as Figure \ref{fig1} suggests. This behaviour could mean that there is a second dynamo (or a string of dynamos)  operating beside the one generating the global, 11-yr long solar cycle. The longitudinal distribution of solar faculae clearly shows that the vast majority of faculae occurred near AL. Faculae with strong magnetic field strength show similar behaviour as to that of the longitudinal distribution of sunspot groups. However, faculae with weak magnetic field strength also show AL-like features. We may therefore provide evidence for the manifestation of the non-axisymmetric solar dynamo at high latitudes, suggesting that the dipolar component of the global magnetic field plays a crucial rule on the behaviour of AL.

\noindent
\textbf{Acknowledgement:} RE is grateful to STFC, RE, NG and MBK acknowledge The Royal Society (UK) for the support received. AE is partially supported by HSPF (Hungarian Solar Physics Foundation).

\end{document}